%% file: main.tex
\documentclass[final,5p,times,twocolumn]{elsarticle}

\usepackage{amssymb}

 \biboptions{sort&compress}

\usepackage{float}
\usepackage{subfig}
\usepackage[export]{adjustbox}
\usepackage{placeins}
\usepackage{booktabs}
\newcommand{\head}[1]{\textnormal{\textbf{#1}}}
\usepackage{amsmath,bm}
\usepackage{pdflscape}
\usepackage{url}
\usepackage{textgreek}
\usepackage[final]{microtype}
\usepackage{multirow}

\usepackage[usenames]{color}

\usepackage{adjustbox}

\usepackage{dblfloatfix}

\newcommand{\beginsupplement}{%
        \setcounter{table}{0}
        \renewcommand{\thetable}{S\arabic{table}}%
        \setcounter{figure}{0}
        \renewcommand{\thefigure}{S\arabic{figure}}%
        \setcounter{section}{0}
        \renewcommand{\thesection}{S\arabic{section}}
     }

\journal{Additive Manufacturing}

\makeatletter
\def\ps@pprintTitle{%
 \let\@oddhead\@empty
 \let\@evenhead\@empty
 \def\@oddfoot{}%
 \let\@evenfoot\@oddfoot}
\makeatother

\begin{document}

\begin{frontmatter}

\title{Radiometric Temperature Measurement for Metal Additive Manufacturing via Temperature Emissivity Separation}

\author[MIT]{Ryan~W.~Penny\corref{cor1}}
\ead{rpenny@mit.edu}
\author[MIT]{A.~John~Hart\corref{cor1}}
\ead{ajhart@mit.edu}

\address[MIT]{Department of Mechanical Engineering, Massachusetts Institute of Technology, 77 Massachusetts Avenue, Cambridge, 02139, MA, USA}

\cortext[cor1]{Corresponding author}

\begin{abstract}

Emission of blackbody emission from the meltpool and surrounding area in laser powder bed fusion (LPBF) makes this process visible to a range of optical monitoring instruments intended for online process and quality control.  Yet, these instruments have yet to prove capable of reliably detecting the finest flaws that influence LPBF component mechanical performance, limiting their adoption. One hindrance lies in interpreting measurements of radiance as temperature, despite the physical link between these variables being readily understood as a combination of Plank's Law and spectral emissivity.  Uncertainty in spectral emissivity arises as it is nearly impossible to predict and can be a strong function of wavelength; in turn, this manifests uncertainty in estimated temperatures and thereby obscures the LPBF process dynamics that indicate component defects.  This paper presents temperature emissivity separation (TES) as a method for accurate retrieval of optically-measured temperatures in LPBF.  TES simultaneously calculates both temperature and spectral emissivity from spectrally-resolved radiance measurements and, as the latter term is effectively measured, more accurate process temperatures result.  Using a bespoke imaging spectrometer integrated with an LPBF testbed to evaluate this approach, three basic TES algorithms are compared in a validation experiment that demonstrates retrieval of temperatures accurate to $\pm 28$~K over a $1000$~K range.  A second investigation proves industrial feasibility through fabrication of an LPBF test artifact.  Temperature data are used to study the evolution of fusion process boundary conditions, including a decrease in cooling rate as layerwise printing proceeds. A provisional correlation of temperature fields to component porosity assessed by 3D computed tomography demonstrates in situ optical detection of micron-scale porous defects in LPBF.

\end{abstract}

\begin{keyword}
additive manufacturing \sep process monitoring \sep spectroscopy \sep quality control \sep infrared
\end{keyword}

\end{frontmatter}

\input{introduction}
\input{methods}
\input{results}
\input{conclusion}

\section{Acknowledgments}

This study was funded by Honeywell Federal Manufacturing \& Technologies (FM\&T) and by a MathWorks MIT Mechanical Engineering Fellowship (to R.W.P.).

\bibliographystyle{elsarticle-num}
\bibliography{refs09_27_2021}

\section{Author Contributions}

R.W.P. and A.J.H. conceptualized ADM, designed the experiments, and drafted the manuscript. R.W.P designed and constructed the imaging spectrometer, performed the experiments, and analyzed the results.  A.J.H. supervised the project, including guiding data analysis, obtaining funding, and securing access to project resources, and revised the manuscript. Both authors read and approved the final manuscript.

\input{supplement}

\end{document}

%% file: introduction.tex
\section{Introduction}

Laser powder bed fusion (LPBF) is the premier process for additive manufacturing of metal components, including those where intricate geometries bring performance advantages~\cite{Wohlers2024}.  However, achieving these benefits in practice is challenged by microscopic porous defects that arise intrinsically, even in the highest quality components~\cite{Thomas2016, Kasperovich2016, Zhang2017,King2014, Calta2018, Chivel2013}.  Pores reduce mechanical properties, particularly fatigue life~\cite{Carlton2016, Nadot2019, Yadollahi2018, Poulin2018}, limiting the use of LPBF components in cyclically-loaded applications~\cite{Blakey2021, Gruber2023, Nadot2019}.  LPBF process parameters that produce high density 
often prove specific to a machine, geometry, print material, and arrangement of other components on the build platform, among other restricting factors~\cite{Frazier2014,Seifi2016,Oliveira2020}.  Even with carefully chosen process parameters, time consuming and costly post-process inspection remains necessary to certify the absence of such defects or, equivalently, bound expected fatigue life~\cite{Seifi2016,Chua2024}.  

In-process optical sensing is commonly presented as a potential solution to these needs, both in deterministically guiding the selection of process parameters and providing for online component qualification~\cite{Everton2016, Grasso2016, McCann2021, Chua2024}.  However, even the most sensitive instruments described in the literature have yet to prove capable of detecting the finest pores that can influence mechanical properties~\cite{Everton2016, Grasso2016, McCann2021, Chua2024, Penny2024_ADM}.  One often-identified impediment to improving the fidelity of optical process measurements is difficulty in interpreting measurements of radiance, or how brightly material glows, as temperature~\cite{Rodriguez2015, Everton2016}.  Inherent spatial and temporal variation in the efficiency of light emission in the region where the powder is being fused, specifically embodied as the optical property of spectral emissivity, prevents accurate temperature estimation via calibration and thereby obscures fusion process dynamics.  

Most optical instruments used in LPBF rely on blackbody radiation to interrogate thermal emission.  A blackbody is an object perfectly efficient at absorbing and emitting radiation, and glows with a spectrum determined solely by its temperature.  How brightly the material glows is called radiance as given by Planck's Law~\cite{Planck1900A, Planck1900B, Planck1901}, 
\begin{equation}
    \label{eq:SpectralPlancksLaw}
    B_\lambda\left(T \right) = \dfrac{2hc^2}{\lambda^5}\dfrac{1}{e^{\dfrac{hc}{\lambda k_B T}}-1},
\end{equation}
for a wavelength $\lambda$, where the additional variables include temperature ($T$), the speed of light ($c$), and Boltzmann's constant ($k_B$).  By inspection, this equation is readily invertible for temperature provided a measurement of $B_\lambda$.  Naturally, no object meets this idealization, let alone the range of optical properties embodied by powdered feedstock, molten material, and the solidified component in LPBF.  It is therefore customary to adapt Eq.~\ref{eq:SpectralPlancksLaw} per
\begin{equation}
\label{eq:SpectralEmissivityDef}
L_\lambda\left(T \right) = \varepsilon_\lambda B_\lambda(T)
\end{equation}
where $\varepsilon_\lambda$ is spectral emissivity that conceptually acts as an efficiency term as a function of wavelength.  Emissivity values near zero indicate highly reflective surfaces and emissivity values near one indicate absorptive (dark) and often rough surfaces.  If $\varepsilon_\lambda$ is known, Eq.~\ref{eq:SpectralEmissivityDef} remains invertible for temperature.

\begin{figure}[htbp]
\begin{center}
	{
	\includegraphics[trim = {2.65in 1.3in 2.7in 1.65in}, clip, scale=1, keepaspectratio=true]{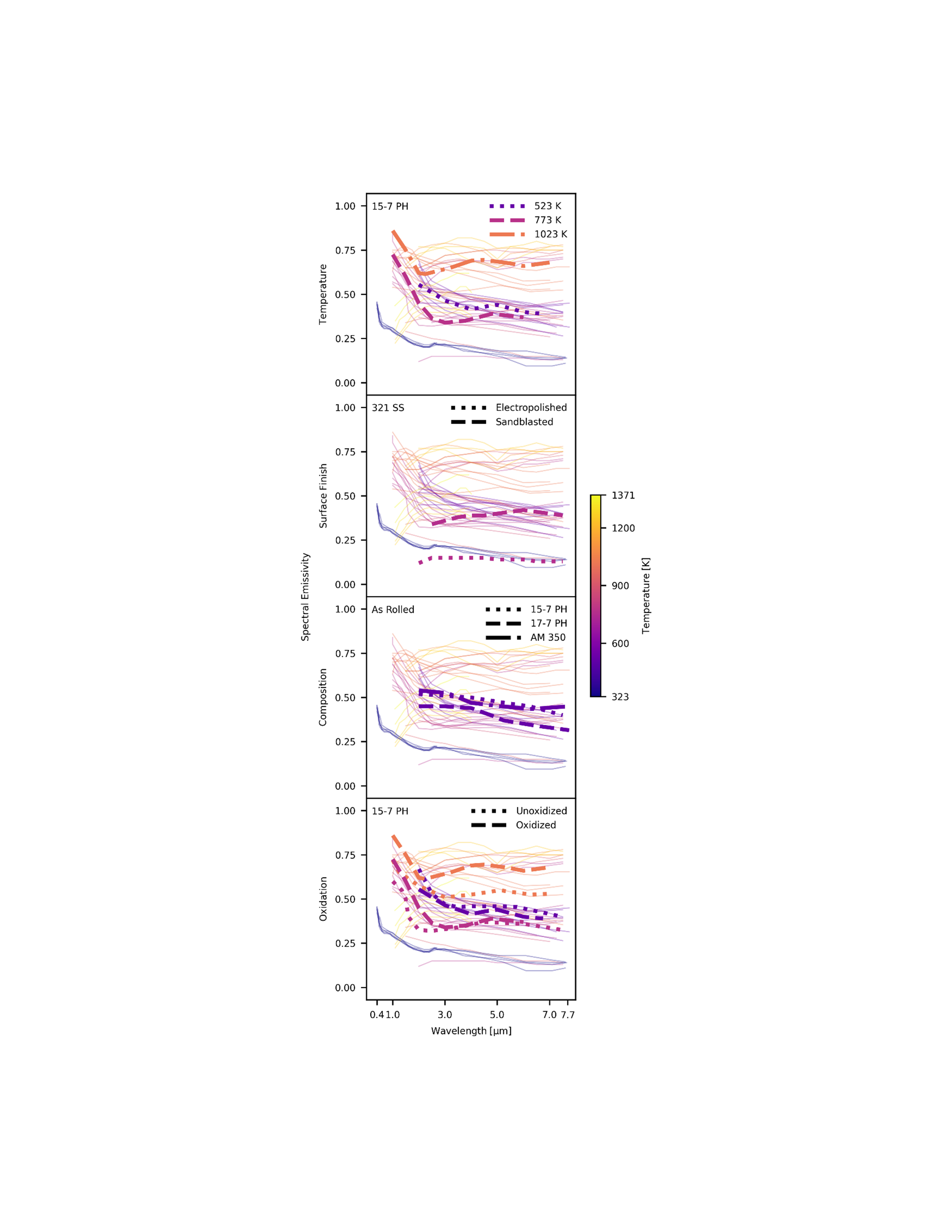}
	}
\end{center}
\vspace{-11pt}
\caption{Emissivity measurements of stainless steel coupons.  All curves are reproduced in all panels; however, bold curves isolate the effect of the variable identified by each subtitle.  Data are adapted from~\cite{Touloukian1970}.}
\label{fig:SpectralEmissivity}
\end{figure}

With this understanding, two approaches can be described for optical process monitoring of LPBF. The first is to simply work with camera data directly, taking emissivity to be an unknown constant factor, and using relative differences in measured radiance to provide insight to the fusion process~\cite{Zhirnov2015, Bartlett2018, Mohr2020}.  Second, and much more common where absolute temperature estimates are desired, is to employ an emissivity calibration scheme.  In this direction, Chivel and Sumurov report first calibrating a silicon CCD sensor operating at visible wavelengths with a tungsten halogen lamp, then use textbook emissivity values to reinterpret the images as temperature fields~\cite{Chivel2011}.  This instrument was primarily used to show an increase in temperature when fusing overhanging features in~\cite{Chivel2011} and for a detailed study of Rayleigh-Taylor instability of the meltpool in~\cite{Chivel2013}.  Lane and colleagues take the same approach, assuming a spectrally-constant emissivity of $0.5$ for Inconel 625, after direct calibration of sensor readings to the temperature of a blackbody~\cite{Lane2016}.  A straightforward alternative is presented in~\cite{Doubenskaia2015}, where a signal level (in counts) is correlated to melting temperature by measuring the width of single laser tracks.  In the $3-5$~\textmu m range, the authors conclude the emissivity of 304 stainless steel is $0.4$ and measure $0.53$ for 316L under typical process conditions.  Another attempt at calibration of a wideband sensor is recounted in~\cite{Williams2019}, wherein an emissivity scheduling approach is investigated.  For solidified material, a calibration target structure is fabricated via LPBF, fitted with a cartridge heater, and instrumented with a thermocouple.  Using this apparatus, an empirical relationship between target temperature and sensor value is recorded (cf.~\cite{Heigel2020} where coefficients of a similar relation are fit to an assumed functional form).  Separately, a second calibration determines a similar relationship for powder.  It is found that an underlying surface with powder has a considerably higher emissivity than the original surface.  Inter-layer cooling time is measured during the recoating step, and this is proposed as process signature for process control.  Rodriguez and coworkers apply a strikingly similar approach using a LWIR camera to study electron beam melting~\cite{Rodriguez2015}.  The emissivity of fused Ti-6Al-4V is observed to be $0.26$ versus $0.50$ for the powder at this wavelength range, as determined by comparison of radiometrically measured temperatures with thermocouple readings.  The authors focus on a restricted temperature range around $760^\circ$C during layer preheating before fusion.  They find that small differences in the temperature field correspond to the energy necessary to melt the material and can be trimmed out during the fusion step.  Additionally, their error analysis indicates that a part surface with an uncorrected radiometric temperature measurement of $1038^\circ$C can actually only be at a temperature of $672^\circ$C against a thermocouple baseline, clearly illustrating the perils of ignoring emissivity when determining absolute temperatures.  It is also noted by the authors that measuring temperatures specific to the actual fusion step presents a more formidable future challenge.

In view of this range of reported emissivity values, it is nearly impossible to know what $\varepsilon_\lambda$ is in the context of LPBF.  This difficulty is further exemplified in Fig.~\ref{fig:SpectralEmissivity}, which presents a selection of spectral emissivity measurements for stainless steel coupons of varying temperature, surface finish, composition, and oxidation level (underlying data are provided in~\cite{Touloukian1970}).  Even the temperature dependence of $\varepsilon_\lambda$ alone eliminates any assumption of a constant emissivity or known spectral emissivity as suitable to the problem at hand.  Moreover, the data presented in this figure are for solid and flat (to the surface finish) specimens; this is not the case in LPBF where material can be powder, liquid, or solid, each with their own unique radiative character.  

More advanced instruments employ two-color pyrometry that specifically aims to estimate temperature in a manner that cancels the effect of unknown emissivity.  This is achieved using radiance measurements centered at two different wavelengths, and an assumption of spectrally-constant emissivity (sometimes called graybody behavior).  The ratio
\begin{equation}
    \dfrac{L_{\lambda 1}\left(T \right)}{L_{\lambda 2}\left(T \right)} = \dfrac{\varepsilon B_{\lambda 1}(T)}{\varepsilon B_{\lambda 2}(T)}
\end{equation}
is constructed.  Therein, the assumption of constant emissivity allows for it to be canceled, making the ratio independent of emissivity.  The final step is inverting this relation for temperature; however, this is complicated by the functional form of Eq.~\ref{eq:SpectralPlancksLaw}.  Rather, the Wien approximation\footnote{Development of this equation predates Planck's solution, though in an attempt to solve the same problem on purely thermodynamic grounds~\cite{Wien1897}.  As such, it is not identified as an approximation in the original work.},
\begin{equation}
\label{eq:SpectralWein}
	B_\lambda(T) = \dfrac{2hc^2}{\lambda^5}e^{-\dfrac{hc}{\lambda k_B T}},
\end{equation}
is most commonly used, wherein all operations, variables, and constants are used in their customary manner~\cite{Wien1897}.  Once $T$ is determined, it is, of course, possible to use Eq.~\ref{eq:SpectralWein} and either of $L_{\lambda_1}$ and $L_{\lambda_2}$ to deduce $\varepsilon$.    It should be noted from simultaneous inspection of Eqs.~\ref{eq:SpectralPlancksLaw} and~\ref{eq:SpectralWein} that these relations are equivalent only in the limit $\dfrac{hc}{\lambda} >> k_BT$.  In other words, the consequence of using the Wein Approximation is a restriction on the wavelengths that yield accurate temperature estimates.  An exemplary computation for a temperature of $2000$~K, germane to LPBF of many alloys, suggests the bands should be chosen at about $720$~nm or shorter.

The efficacy of this approach to measuring metals near their melting point is rigorously investigated in~\cite{Monier2017}.  Therein, the precision of two-color pyrometry is shown to depend on a number of factors, including the wavelengths of the band centers, band widths, and the temperature of the object.  Critically, the authors explain that the design of these instruments encounter a critical engineering decision.  On one hand, placing the band centers farther apart gives more contrast between the channels and can thereby increase the fidelity of temperature measurement, where error is proportional to
\begin{equation}
\dfrac{T\lambda_1\lambda_2}{\dfrac{hc}{k_B}\left( \lambda_2 - \lambda_1 \right)}.
\end{equation}
On the other hand, this benefit lasts only to the extent the greybody approximation still holds, and this is decreasingly plausible as the wavelength spacing is increased.  Monier presents experiments that evaluate pyrometers operating with different combinations of band centers, using the melting points of pure metals as a benchmark, and find that temperature errors between $5\%$ and $12$\% are typical in this application~\cite{Monier2017}.  Nonetheless, Hooper and colleagues apply a two color pyrometer to LPBF in~\cite{Hooper2018}.  In addition to reiterating the prior limitation, they additionally find that wavelength choice is further limited by coatings on other optics if the path is shared with the laser.  While no quantifiable metrics of temperature fidelity are noted, a high sensitivity to image noise is reported.  Using a scanned pyrometer, Vecchiato and colleagues recover optical temperatures sufficient to relate cooling rates and thermal gradients to component microstructure~\cite{Vecchiato2020}.  Finally, a commercial two-color instrument is proven capable of resolving pores of $70$~\textmu m diameter and larger in~\cite{Mitchell2020}, making it among the most sensitive instruments applied to LPBF quality control~\cite{Penny2024_ADM}.  While temperature-resolved process signatures are shown to be highly effective at identifying likely-porous regions of LPBF parts, impacts of the deviation of stainless steel from greybody behavior process signature efficacy are not investigated~\cite{Mitchell2020}.  Further, this assumption is made over a temperature range of $1100-2800^\circ$C and wavelength range spanning $750-900$~nm.  As seen in Fig.~\ref{fig:SpectralEmissivity}, despite most curves stopping just shy of this range, spectral emissivity can be a strong function of wavelength at around $1$~\textmu m.

This work investigates temperature-emissivity separation (TES) as a method for improving the accuracy of optical temperature measurements of LPBF.  This method simultaneously estimates both temperature and spectral emissivity using spectrally-resolved data; thus, the temperature estimates are less sensitive to this difficult-to-anticipate optical property~\cite{Watson1992,Li2019}.  The origins of TES lie in spacebourne and airborne remote sensing in the long wave infrared, with original applications in geology, meteorology, pollution monitoring, an surveillance (a catalog of platforms and experiments is available in~\cite{Kramer2002}), where spectral emissivity usually is the signal of interest and it is unknown temperature which complicates its retrieval.  The problem faced here is the opposite side of the same coin, namely that temperature is the signal of interest and its retrieval is complicated by unknown emissivity.  Adaption of TES to the situation at hand requires some explanation of the native application of these algorithms, particularly as they often include features that are unnecessary in the present application, and the reader is referred to the landmark text by Chandrasekhar~\cite{Chandrasekhar1960} for this purpose.

TES for retrieving accurate process temperatures in the context of LPBF is enabled by integrating a custom-built imaging spectrometer with an LPBF testbed, as schematically illustrated in Fig.~\ref{fig:Fig2}a.  With this instrumentation, point melting experiments are first performed to prove the accuracy of the TES approach.  Laser-induced melting of a selection of pure metals and alloys are observed and melting points are retrieved via TES from the spectrometer data.  These are compared to their textbook values, proving $\pm 28$~K fidelity over more than a $1000$~K range.  Second, fabrication of an LPBF test artifact demonstrates this technology in a production-relevant environment.  Results include quantification of cooling rate, which is observed to decrease as the print proceeds and heat accumulates, among other indications of evolving boundary conditions.  Finally, demonstrating this approach for quality control of LPBF, an exploratory study shows how anomalous time-temperature history can predict the presence of micro-CT-resolved internal pores as small as $\geq 4.3$~\textmu m diameter.

%% file: methods.tex
\section{Methods}
\label{sec:Methods}

\begin{figure*}[htbp]
\begin{center}
	{
	\includegraphics[trim = {1in 3.4in 1in 3.25in}, clip, scale=1, keepaspectratio=true]{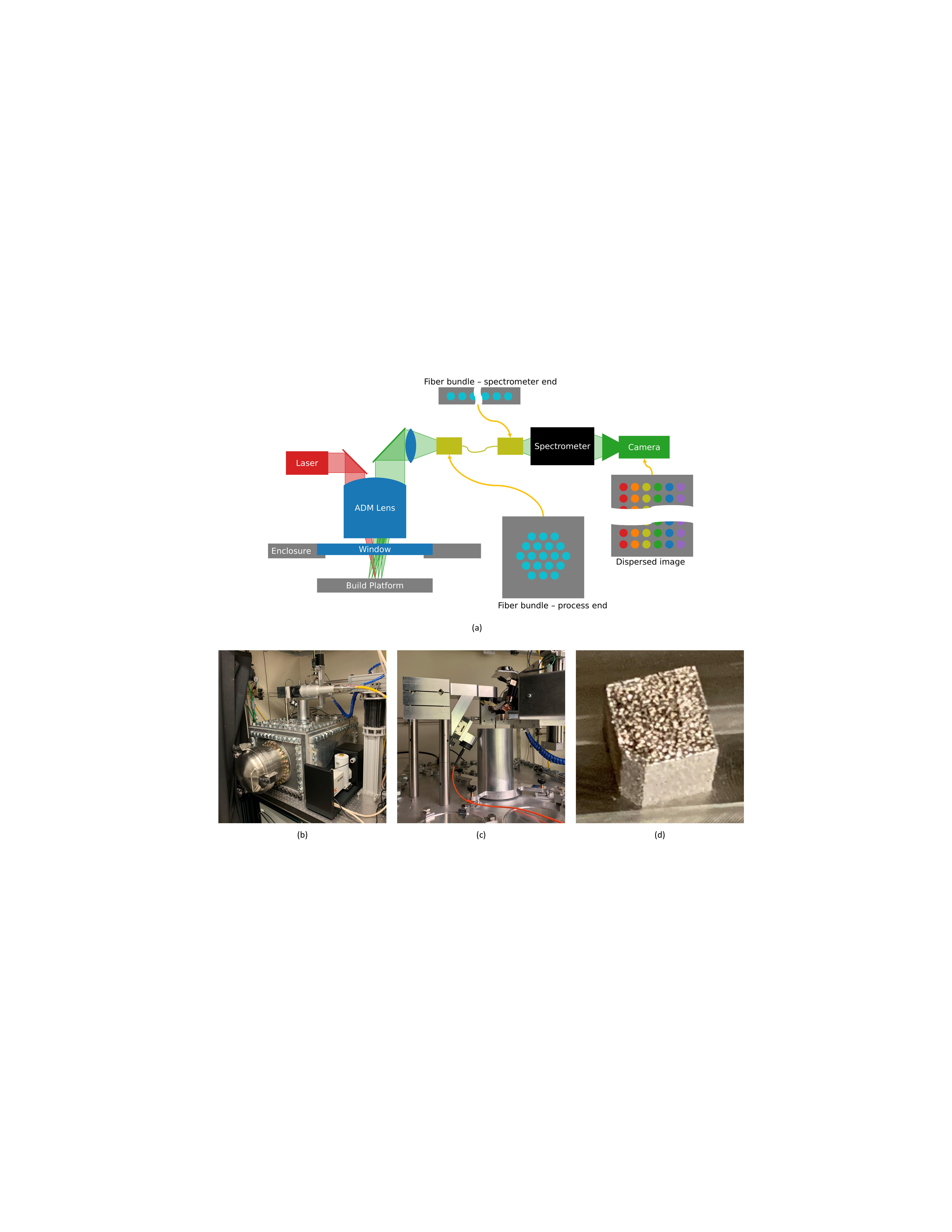}
	}
\end{center}
\caption{(a) Schematic illustration of imaging spectrometer coupled to the LPBF testbed using aperture division multiplexing.  (b) Experimental apparatus, showing the LPBF testbed, laser delivery boom, ADM lens, and imaging spectrometer.  (c)  Detailed view of the optical relay, including the ADM lens and ADM lens interface.  (d) LPBF test artifact.}
\label{fig:Fig2}
\end{figure*}

Figure~\ref{fig:Fig2}a schematically illustrates the key components of the apparatus, constructed around the custom LPBF testbed shown in Fig.~\ref{fig:Fig2}b.  Uniquely, the build platform is viewed simultaneously by the process laser and spectrometer by aperture division multiplexing (ADM)~\cite{Penny2024_ADM}.  One portion of the ADM optic focuses the laser light and a second portion forms one half of an imaging relay.  The opposite half of the relay is a separate optical assembly that images the build platform onto one end of a fiber bundle with a close-pack hexagonal array of optical fibers, which are rearranged into a linear array at the opposite end of the fiber bundle.  An imaging spectrometer disperses the light emerging from each fiber at the linear end of the bundle, creating an array of spectra that are recorded by a camera.  Spectrometer calibration and temperature-emissivity separation are performed in post-processing.

\subsection{LPBF Testbed}

All experiments reported here use the custom-built LPBF testbed visible in Fig.~\ref{fig:Fig2}b. The testbed is fully described in~\cite{Penny2024_Thesis, Kutschke2023} and builds upon prior work described by~\cite{Baker2017, Gibbs2018, Griggs2021}.  This machine replicates the features of commercial LPBF equipment, including a piston-fed powder supply, mechanized recoater with a compliant blade, inert atmospheric control, and directed gas flow for spatter removal.  Uniquely, the distance from the build plane to the inner surface of the environmental window is only $60$~mm.  This provides a short working distance for laser delivery and imaging optics, including the ADM lens.  Figure~\ref{fig:Fig2}b also shows a scan head affixed to a boom above the printer enclosure.  This assembly comprises a $500$~W, $1.075$~\textmu m single-mode fiber laser (redPOWER qube, SPI) and a $250$~mm collimator (Coherent PN 106402X01), which generate a $20$~mm diameter beam.  This beam is first redirected by a turning mirror (Thorlabs PN NB2-K14) before entering a galvanometer mirror set (Thorlabs PN QS20XY).

The laser has a minimum rated power of $50$~W and this power is too high to perform the point melting experiments described below as the spot is stationary.  Accordingly, for those experiments the turning mirror is substituted with a beam splitting optic (Thorlabs PN BSS17) that reduces the power reflected to the galvanometers by $\approx 70$\%. The unused portion of the beam, which is transmitted through this optic, is directed into a beam dump (Ophir PN BDFL500).

\subsection{Imaging Spectrometer}
\label{sec:MethodsSpectrometer}

To provide the spectrally and spatially resolved data necessary for temperature emissivity separation, the imaging spectrometer shown in Fig.~\ref{fig:FigS1}a is used (a labeled CAD model is provided in Fig.~\ref{fig:FigS1}b and it is shown as-integrated in Fig.~\ref{fig:Fig2}b).  The instrument is coupled to the rest of the apparatus with a custom $2$~m long fiber bundle (Berkshire Photonics).  It comprises $19$ individual low-OH fibers with a $100$~\textmu m core, $125$~\textmu m cladding, and $0.22$~NA.  At the spectrometer end of the bundle, the fibers are stacked in a line within a $10$~mm diameter ferrule.  This is positioned at the object plane, without a slit.  The spectrometer itself is a half-crossed Czerny-Turner design operating at $f/2.3$.  One end of the operating band is set at $1.2$~\textmu m via an interference filter (Thorlabs PN FELH1200). Instrument response rolls off at $2.4$~\textmu m as a consequence of transmission through the optical fiber bundle.  The spectrometer design and calibration are fully explained in \S~\ref{sec:SupSpectrometer} and suggest a limiting spectral resolution of $40$~nm ($30$~bands).  Finally, a high-speed MWIR camera (IRCameras IRC806HS) is used as a detector, equipped with an $f/2.3$ cold stop but not a cold filter.

\subsection{ADM Lens}

Figure~\ref{fig:Fig2}c shows both the laser and spectrometer are interfaced with the LPBF testbed by a single optic called the ADM (aperture division multiplexing) lens; the reader is referred to~\cite{Penny2022_ADM} for a general overview of this technology and to~\cite{Penny2024_ADM, Penny2024_Thesis} for complete details of this specific implementation.  To summarize, this nominally $125$~mm focal length optic provides two optical paths.  Collimated laser light enters one optical path, after redirection by the galvanometer set, that is designed to focus the laser to an $\approx 77$~\textmu m diameter (D86) spot with the theoretical focus lying below the build plane.  The second optical path uses a different portion of the same optic, albeit in the reverse direction, as one half of an imaging relay for process monitoring.  This path is nominally designed to function with the experimental setup described in~\cite{Penny2024_ADM}, where the relay is completed with a $50$~mm, $f/2.3$ optic that defines the limiting aperture of the system.  For the present spectrometer configuration, also shown in Fig.~\ref{fig:Fig2}c, the relay is completed by a $136.1$~mm off-axis parabolic mirror (Edmund Optics PN 35-614).  This optic focuses light exiting the ADM lens onto the testbed-end of the fiber bundle, where the fibers are disposed in a close-pack hexagonal array in a standard SMA connector.  The ratio of focal lengths suggests that the image of the build plate on the process end of the fiber bundle is magnified by a factor of $1.09$.  Optical throughput is limited by the finite dimensions of this optical path through the ADM lens, namely by the D-shaped element at the top that forms the limiting aperture of the ADM optic, and so the fibers are slightly underfilled.  A final consequence of this arrangement is that reflected laser light comes to a focus near the testbed-end of the fiber bundle.  To prevent this from causing damage, a reflective longpass filter (Thorlabs PN FELH1200) is placed between the parabolic mirror and the fiber bundle tip.

\subsection{Temperature Emissivity Separation}
\label{sec:MethodsTES}

TES resolves the present obstacle, which is made clear by considering it as a system of equations.  When imaging in a single band, one measurement is available to solve for temperature and emissivity.  Two-color pyrometry makes two measurements available, yet it is necessary to solve for temperature and two emissivities (if greybody behavior is not assumed).  This line of logic continues for an arbitrary number of measurements,
\begin{equation}
\begin{alignedat}{2}
L_1 &= \varepsilon_1 B_1(T)\\
L_2 &= \varepsilon_2 B_2(T)\\
\vdots\; &= \phantom{\varepsilon_1 B}\vdots\\
L_n &= \varepsilon_n B_n(T),
\end{alignedat}
\end{equation}
where radiance measurements at $n$ wavelengths underdetermine $n$ emissivities and $1$ temperature.  In other words, spectrally-resolved data do not independently provide a solution for emissivity correction.  Thus, the objective of TES is to provide a sufficient number of equations to solve this system, in view of at least one spectrally-resolved radiance measurement and through minimally-limiting assumptions of emissivity that can match the optical character of real materials~\cite{Watson1992_Ratios,Borel1998,Gillespie1998,Li1999,Borel2003,Boonmee2006, Bioucas2012,Xu2013,Qian2016}.


The three TES methods used in this study break the dependence between the number of spectral measurements and number of variables to be fit by assuming at least some degree of invariance along one axis of the spatial-spectral-temporal (three-axis) datasets generated. To their advantage, these are some of the least limiting TES methods, while being straightforward to implement.  Specifically, two methods make no assumptions about the shape or mean of the emissivity values retrieved and the third only requires that emissivity be the same for a pair of spectral bins among a larger number measured.

The first selected approach to TES is described by Watson in~\cite{Watson1992}, and is now commonly called the Two Temperature Method (see~\cite{Peres2006} for an updated derivative).  Rather than be concerned with recovering temperatures, however, Watson is originally interested in measuring accurate emissivities from spectrally-resolved satellite image data for geographical study.  Watson's approach is to image a region of interest with an $n$ band imaging spectrometer during the day, when the sun has warmed the ground surface, then return at night and measure the same scene again when it has cooled to a second, lower temperature.  Assuming that the emissivity did not appreciably change between the two measurements, i.e., it is only a weak function of temperature, only $n$ spectral emissivity values need to be determined.  Thus, $2n$ measurements are available to fit $n+2$ unknowns ($n$ emissivities and two temperatures) in an over-determined system of equations, and these variables are determined with no prior knowledge.  In the context of LPBF, it is quite easy to measure the same point at a wide range of temperatures and so this approach can be well-suited.

Spatial correlations are also commonly leveraged in TES~\cite{Barducci2004,Xu2013} and at a simple level can be applied analogously to the two temperature method.  In this case, one collects a spectrally-resolved radiance measurements in $n$ bands for each of a first and a second location, and assumes that emissivity is constant between them.  Then, and same as before, $2n$ measurements are available to fit $n+2$ unknowns ($n$ emissivities and $2$ temperatures).  Selection of the locations is fixed in the implementation used here, where a neighboring channel has been chosen.  However, it should be noted that it is possible to pick the second location dynamically.  For example, the spectral angle (or any other metric of spectral distance) can be computed between the spectral radiance measurement at first location and each of the measurements of the 8 surrounding pixels.  The surrounding pixel with the most similar spectrum is then used as the second location for TES.

A third and final approach assumes invariance along the spectral dimension of such a dataset and is known as the greybody method, as described by Barducci and Pippi~\cite{Barducci1996}.  The greybody method operates on only a single measurement in $n$ bands.  For two bands, which can be chosen arbitrarily or in view of prior knowledge, the emissivity is set to be equal.  Then, $n-1$ emissivities and $1$ temperature ($n$ parameters in total) are fit to the $n$ radiance measurements.  This can be a good fit for instruments with fine spectral resolution.  The emissivity for the two longest-wavelength bands are set equal here.

In all cases, temperature and spectral emissivity values are fit using the minimize function of Sicpy.optimize~\cite{Scipy2020} that is configured to use the Nelder-Mead method.  The optimization is initialized using a temperature of $1700$~K and a constant spectral emissivity of $0.2$. No regularization of the optimization or post-fit filtering are applied, although this may bring substantial benefits to future work.

\subsection{Point Melting Experiments}
\label{sec:MethodsPoint}

Point melting experiments are performed to quantify temperature measurement accuracy, using a strategy similar to that employed in~\cite{Monier2017}.  For this purpose, the laser in the LPBF testbed is used to heat a point on a metal specimen that is also monitored with the spectrometer.  Pure metals demonstrate consistent heating as the experiment proceeds, followed by a pause in temperature change as the latent heat of fusion is overcome at the melting point (i.e., a meltpool forms in the region observed), and then resume increasing in temperature once the region is liquid.  Comparing the optically-measured temperature at the pause to the textbook values provides a means for assessing instrument accuracy.  Melting events are identified via a filtered derivative.  In pure-metal experiments, melting is taken to begin at the first zero crossing and end at the last zero crossing of the filtered derivative.  The average temperature between these times is taken to be the measured melting point and the standard deviation is also computed to understand measurement uncertainty.  Alloys are considered similarly and, while their melting behavior is more complex, their liquidus points are also observed and compared to textbook values.  Liquidus is identified by a significant inflection point, again made clear via filtered derivative of the temperature data.

These experiments begin by placing a metal specimen on the build platform of the LPBF machine and adjusting the position of the build platform such that the top surface of the metal specimen lies in the build plane.  Next, the enclosure is purged to $\leq 400$~PPM of oxygen using argon (Linde, 99.999\%), as measured via MTI W1000 trace oxygen sensor.  Operating the laser at a low power, the turning mirror is adjusted to maximize the net optical power recorded in the central fiber of the hexagonal pattern (i.e., the laser spot is made co-incident with the center of the field of view of the spectrometer).  Then, a series of trials determine laser parameters that induce a clearly-visible melting event in the spectrometer data, where laser power is applied as a ramp over $4$~s in all experiments.  Only the starting and ending laser powers are adjusted, which must be tuned for the optical and thermal properties of each specimen individually.  The LPBF testbed enclosure is then opened to reposition the metal specimen, such that a fresh location on its surface is placed at the at the center of the field of view.  Then, the enclosure purged to $\leq 100$~PPM of oxygen over approximately $45$~minutes, and laser melting and temperature retrieval are performed for the final time.  This two-step protocol is specifically used to limit meltpool oxidation, which is significant in view of the low volume of hot material and extended experiment duration.

Using this method, the selection of pure metals and alloys in Table~\ref{table:SpectralMeltingMetals} are used as benchmark materials; melting point and liquidus data are taken from the compilation in~\cite{Valencia2008}.  In all, they are chosen to provide a range of melting points over roughly a $1000$~K range representative of temperatures in LPBF.  The five pure metals are sourced from Goodfellow as roughly $25 \times 25 \times 3$~mm$^3$ squares except for the copper specimen which is smaller ($10 \times 10 \times 3$~mm$^3$).  Certified alloy specimens are obtained from McMaster-Carr and are cut to similar size with a band saw.  As it has the highest melting point, titanium is used to set the the camera exposure time ($0.125$~ms) to avoid detector saturation; this is held constant when testing the remainder of the specimens.  A $100$~FPS sample rate is used to limit the size of the resulting dataset.

\begin{table*}[bt]
\centering
\footnotesize
\caption[Tabulated properties of melting specimens and laser parameters.]{Tabulated properties of melting specimens and laser parameters.  Melting points listed for alloys are liquidus points.}
\label{table:SpectralMeltingMetals}
\begin{tabular}{@{}*6c@{}}
  
  \toprule[1.5pt]
  \multicolumn{1}{c}{\head{Material}} &
  \multicolumn{2}{c}{\head{Melting Point [K]}}&
  \multicolumn{1}{c}{\head{Nominal Purity [\%]}} &
  \multicolumn{1}{c}{\head{Supplier}} & \multicolumn{1}{c}{\head{PN}}\\
  
   & \head{Reference} & \head{Measured} & & & \\
  
  \cmidrule{1-6}
   
    Ti & 1940 & 1919 & 99.99 & Goodfellow & TI00-FL-000400 \\
    Fe & 1809 & 1816 & 99.99+ & Goodfellow & FE00-FL-000560 \\
    Ni & 1726 & 1715 & 99.99+ & Goodfellow & NI00-FL-000222 \\
    Cu & 1356 & 1331 & 99.99+ & Goodfellow & CU00-FL-000266 \\
    Al & 933 & 905 & 99.999 & Goodfellow & AL00-FL-000360 \\

  \cmidrule{1-6}
    
    Titanium (Ti-6Al-4V) & 1923 & 1912 & - & McMaster-Carr & 9039K24 \\ 
    316 Stainless Steel & 1723 & 1739 & - & McMaster-Carr & 88885K76 \\ 
    Inconel 718 & 1609 & 1602 & - & McMaster-Carr & 1099N606 \\ 
 
  \bottomrule[1.5pt]
\end{tabular}
\end{table*}

\subsection{Build Planning and Printing}

Efficacy of the MWIR imaging spectrometer instrumentation and the data processing approach are shown in an industrially-relevant setting by monitoring the printing of an LPBF test artifact.  Figure~\ref{fig:Fig2}d shows the design: a simple $5\times 5 \times 6$~mm$^3$ cube with a $1\times 1$~mm chamfer added along one vertical edge to track component orientation.  After defining this basic geometry in Solidworks (Dassault Systems), print planning is completed using Autodesk Netfabb software.  This software is used to slice the component to achieve a $30$~\textmu m layer height, then the perimeter of each slice is offset inward by $35$~\textmu m (roughly half of the laser spot diameter).  An infill pattern is generated within this infill region, rotated $67^\circ$ between each layer, where all hatches in a layer are scanned in the same direction using a $30$~\textmu m hatch spacing.  A copy of the slice perimeter is then made after the hatches, and these two items are exported as a .cli file in that order (i.e., the printer is commanded to fuse the infill, then scan the part perimeter).  In turn, the .cli is converted to machine code using a custom Python script that also sets laser power and scan speed.  For the present experiments, the power is set to $100$~W except for layers 100 to 150 that are printed with $90$~W.  A $250$~mm/s scan speed is used at all times. 

To prepare for printing the D86 laser spot size ($77$~\textmu m) is verified by direct imaging per~\cite{Penny2024_ADM}, then the galvanometer null position (defined to be coincident with the center of the component in the machine code) is aligned with with the center of the spectrometer field of view using a process similar to~\S~\ref{sec:MethodsPoint}.  Next, the printer is loaded with 316~stainless steel powder (Carpenter Technology, $15-45$~\textmu m) and ultra-high purity argon (Linde, 99.999\%). Finally, the camera in the spectrometer is configured to sample at $475$~FPS with the same exposure time of $125$~\textmu s used in the point melting experiments.

%% file: results.tex
\section{Results}

Application of TES to LPBF, relying on the custom imaging spectrometer and LPBF testbed described above, is investigated in three parts.  First, a series of point melting experiments on pure metals and alloys are used to understand the accuracy of this hybrid approach.  Second, fluctuations in fusion boundary conditions are quantified using data collected during the fabrication of a LPBF test artifact and are qualitatively related to build quality.  Finally, process signatures extracted from the time-temperature data are proven predictive of micron-scale component porosity.

\subsection{Point Melting Experiments}

\begin{figure*}[htbp]
\begin{center}
	{
	\includegraphics[trim = {0.6in 1.75in 0.8in 2.4in}, clip, scale=1, keepaspectratio=true]{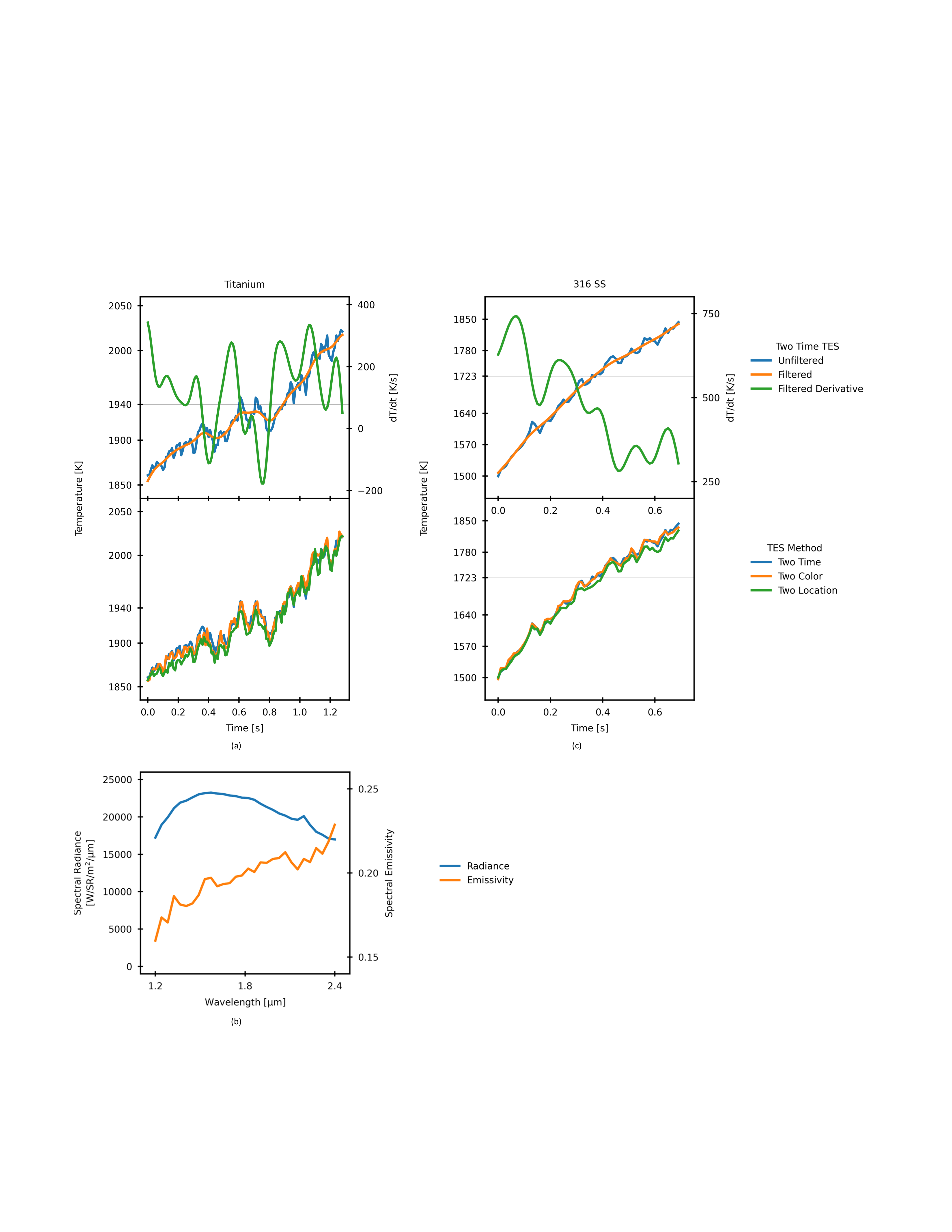}
	}
\end{center}
\caption{(a) Point melting experiment using a pure titanium specimen.  In the top panel, the measured value of $1919$~K is taken to be the average temperature between the first and last zero crossing of the filtered derivative near the melting point.  The reference value of $1940$~K is identified by the light grey lines.  In the bottom panel, all TES methods are shown to give similar results near the melting point.  (b) A typical radiance measurement made with the spectrometer and retrieved emissivity during the melting event observed in (a). 
 (c) Point melting experiment using a 316 SS specimen.  The significance of the panels mirrors (a).  Note that liquidus is associated with an inflection point rather than a pause.}
\label{fig:Fig3}
\end{figure*}

Typical results for point melting experiments are exhibited in Fig.~\ref{fig:Fig3}.  A pure titanium specimen is considered in Fig.~\ref{fig:Fig3}a, wherein the top panel shows that an average temperature of $1919$~K is observed with the two time TES method at the transition from solid to liquid; this compares closely to the expected value of $1940$~K.  During this transition, the measured temperature fluctuates with a standard deviation of $15.2$~K.  In the bottom panel, the two time method is compared to the other TES methods described above, and no significant variation is observed.  Figure~\ref{fig:Fig3}b shows the features that underlying a temperature estimate made in the estimated melting period, including the spectrally-resolved radiance measurement made by the spectrometer as well as the estimated emissivity.

Figure~\ref{fig:Fig3}c shows a similar result for a 316 SS specimen; again, a clear pause in temperature is not evident because this is an alloy.  Nonetheless, the top panel shows a significant inflection point at around $1723$~K which we associate with the liquidus point that is nominally at $1739$~K.  Also like the Ti case, temperature measurements provided by the different TES methods are compared in the bottom panel.

\begin{figure}[htbp]
\begin{center}
	{
	\includegraphics[trim = {2.7in 3.2in 3.2in 2.7in}, clip, scale=1, keepaspectratio=true]{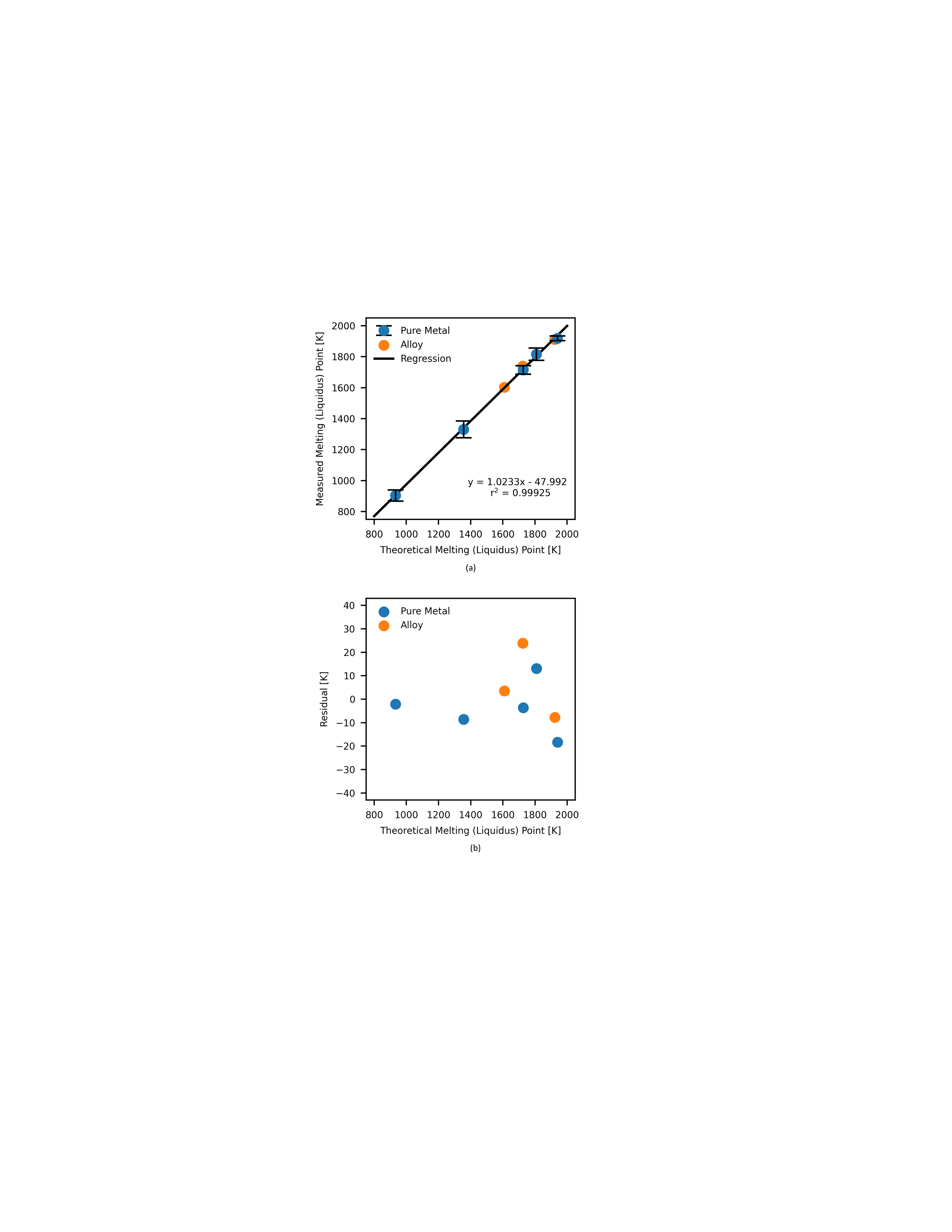}
	}
\end{center}
\caption{Comparison of measured melting and liquidus points against their theoretical values.  (a) A regression indicates a systematic overestimate of $2.33$\%.  (b) Regression residuals from (a).  No further systematic trends are observed.}
\label{fig:Fig4}
\end{figure}

Full results are provided in Table~\ref{table:SpectralMeltingMetals} and show that the instrument is fully capable of resolving temperatures typical of LPBF with a maximum error of $28$~K across the materials studied.  Alternatively, these data are graphically presented in Fig~\ref{fig:Fig4}.  In Fig~\ref{fig:Fig4}a, the observed melting and liquidus points are plotted against their theoretical values.  Error bars are provided for the pure metals and, as described in the methods section, represent the $1$\textsigma{} standard deviation of temperatures spanning the beginning and ending of the observation of melting.  They are largest where low signal levels are recorded, for the copper and aluminum, and shrink as the signal-to-noise ratio improves for meting points at higher temperatures.  No error bars are provided for the alloy measurements as liquidus occurs at a single point in time in these experiments.  Finally, a linear regression is provided to assess the presence of systematic errors in these measurements, which is evident as an overestimate of $2.33$\%.  This is attributed to calibration error and improvements to the calibration process are expected to further improve instrument accuracy.  Examination of the residuals to this regression, provided in Fig.~\ref{fig:Fig4}b, indicates no obvious additional systematic trends.

\subsection{Layerwise Temperature Measurements during LPBF}

\begin{figure*}[htbp]
\begin{center}
	{
	\includegraphics[trim = {1in 3in 1in 1.7in}, clip, scale=1, keepaspectratio=true]{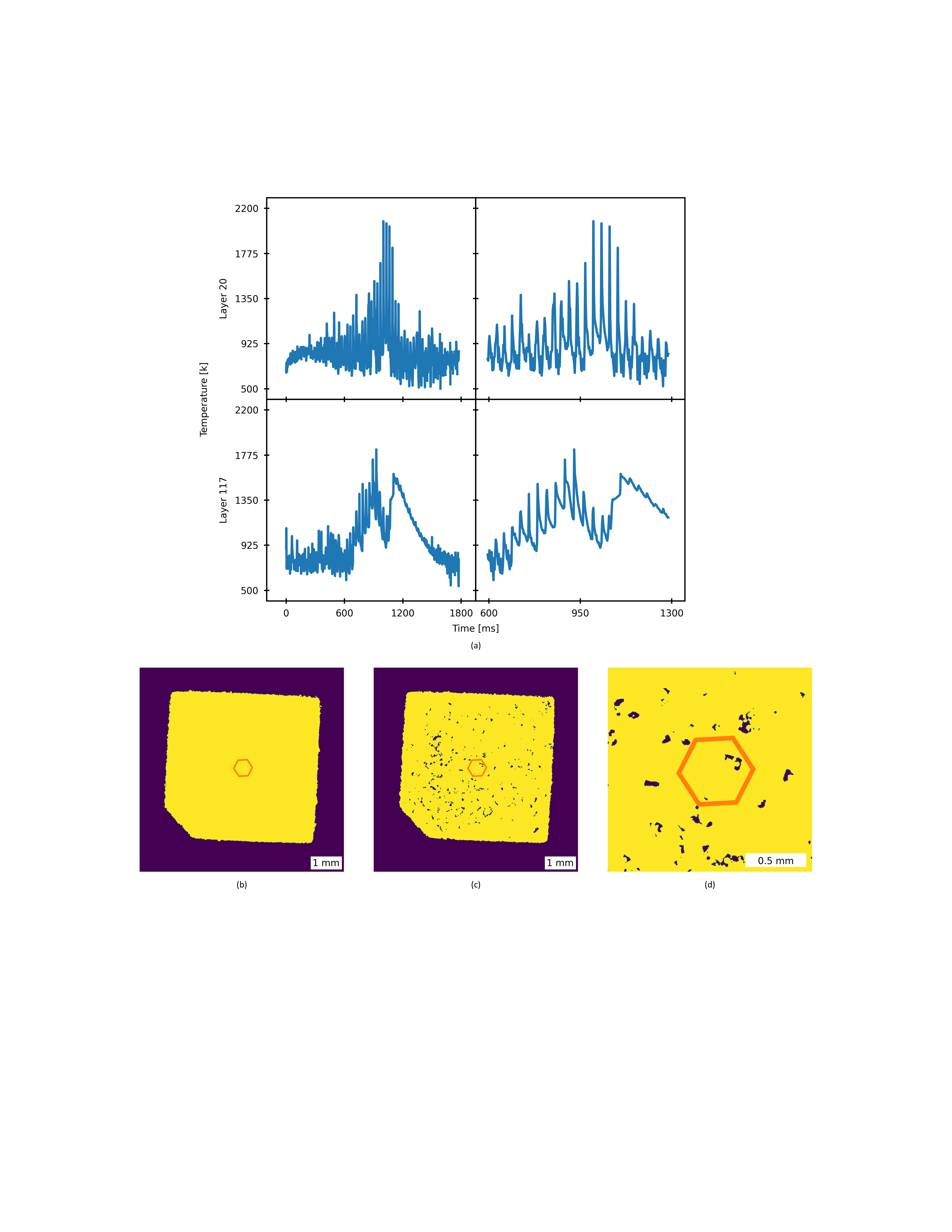}
	}
\end{center}
\caption{(a) Retrieved temperatures for layers 20 and 117 using the two time TES method, where dropouts (temperatures below the instrument noise floor) are removed.  The right panels show an arbitrary portion of these data at a finer time scale.  (b) Micro-CT slice corresponding to layer $20$ where full density is achieved.  Like the remaining figures, the orange hexagon indicates the field of view of the imaging spectrometer.  (c) Micro-CT slice corresponding to layer $117$, showing a significant level of internal porosity.  (d) An enlarged version of (c), wherein a void is found within the field of view of the spectrometer.  The pore shape is hypothesized to be the result of a balling event.}
\label{fig:Fig5}
\end{figure*}

Figure~\ref{fig:Fig5}a shows temperature data, retrieved using the two time TES method operating on data from the center spectrometer channel, for two representative layers during fabrication of the LPBF test artifact.  Layer~20 is characteristic of regions that achieve full density, evidenced via the micro-CT slice in Fig.~\ref{fig:Fig5}b.  An initial increase in temperature is observed at the beginning of layer fusion (around $100$~ms), where thermal energy is conducted from the distant meltpool (i.e. hatching begins a point on the perimeter of the part about 2~mm from the location of observation).  This is the primary cause of the temperature fluctuations leading up to around $950$~ms into the layer, and it increases in magnitude as the scan trajectory approaches the center of the test artifact.  Cyclic heating becomes great enough approximately $1000$~ms into fusion of the layer, where direct laser heating brings the observed material through the liquidus point at $1708$~K no fewer than four times.  Rapid cooling is also observed, where the molten material reaches at temperature of less than $1000$~K between subsequent hatches.  A comparison to the other TES methods can be made with the figures in \S~\ref{sec:SupTES}, which minimal variation much like the point-melting experiments.  The only notable difference is that the two-time method more often recovers temperatures near the noise floor of the instrument, where the optimization fails to converge for the others.

The temperature traces in Layer 20 are contrasted with the traces corresponding to layer~117, where the obvious differences are impaired cooling (particularly from $700$ to $1100$~ms) and a melting event around $1000$~ms where the material takes an extremely long time to cool.  

Inspection of the micro-CT data from these selected layers, plotted in Figs.~\ref{fig:Fig5}b and~\ref{fig:Fig5}c, shows that layer 20 is nearly homogeneous, while layer 117 contains a large number of flaws in the cross section.  The geometric attributes of the flaws are fairly consistent, often comprising large blob-like features of solid material surrounded by a horseshoe-shaped pore.  While these pores sometimes have unfused powder within them, the blobs are far too large to be powder particles themselves and so the underlying mechanism is not lack-of-fusion from insufficient laser power.  Rather, combination of extended cooling and pore shape resolved in micro-CT images suggest that these defects are largely caused by balling; the extended cooling seen in the spectrometer data is molten material that has been pulled into a ball via surface tension and cools slowly on account of its size and limited contact area with the underlying material.  Once formed, this ball continues to disturb the fusion process around it by failing to fully remelt and become consolidated with the surrounding feedstock.  The enlarged area in Fig.~\ref{fig:Fig5}d shows the ball and a pair of severe pores that caused the cooling anomaly in Fig.~\ref{fig:Fig5}a.

\subsubsection{Process Signatures}

Temperature time-series measurements, of which there are $19\, \mathrm{channels} \, \times \, 200 \, \mathrm{layers} \, = \, 3800 \, \mathrm{temperature \, time \, series}$, are re-interpreted as three process signatures (single scalar values) that are hypothesized to indicate porosity.  For example, the simplest is the maximum temperature observed for each time series.  Integral is more complex; it is the sum temperature measurements where the temperature is estimated to exceed $1200$~K.  Finally, cooling rate is estimated.  This begins by identifying peaks in the time series greater than $1200$~K using the signal.find\_peaks routine in SciPy~\cite{Scipy2020} (extra function arguments are specified as a distance of $3$ and prominence of $4$).  Cooling events are generally defined as the $20$ temperature measurements beginning with and following a peak.  However, in cases where a second peak and cooling event occurs within this time, the first event is truncated at the temperature measurement preceding the second peak.  The cooling events are then time-aligned, averaged, and used to fit an equation of the form $a\cdot e^{-bt}$ where $b$ is the cooling rate.

\subsection{Boundary Condition Evolution}

\begin{figure}[htbp]
\begin{center}
	{
	\includegraphics[trim = {1.2in 1.8in 5in 2.1in}, clip, scale=1, keepaspectratio=true]{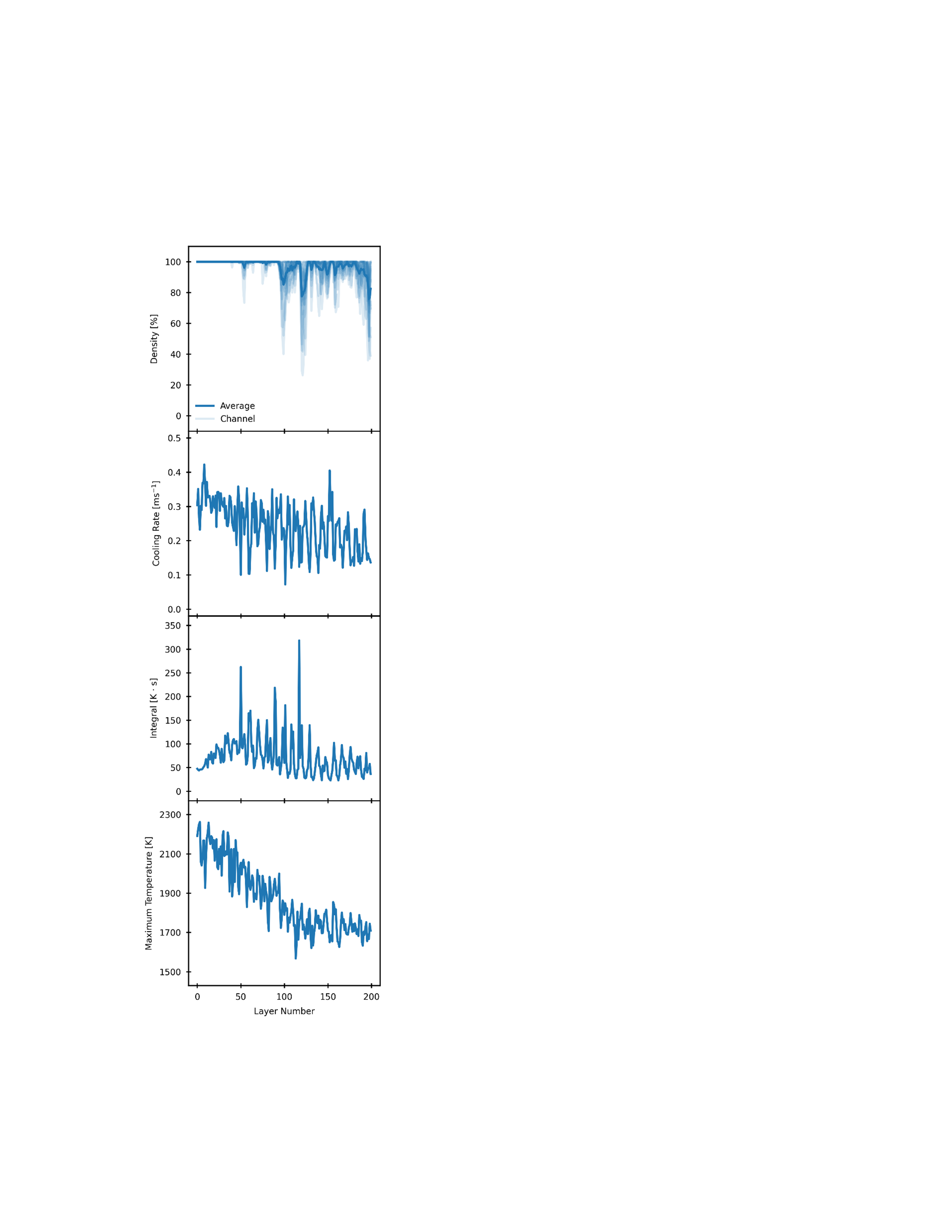}
	}
\end{center}
\caption{Comparison of laser power, test artifact density, and observed boundary conditions.  In the top panel, artifact density and laser power setting are provided for each layer.  The dark blue curve corresponds to the average density over the field of view of the spectrometer and light blue curves correspond to each of the spectrometer channels.  Note that component density falls around layer $100$, where the laser power is decreased from $100$~W to $90$~W.  In the panel below, cooling rate is plotted versus layer number.  Cooling rate initially decreases as the print proceeds and stabilizes around layer $50$.  Lower still is the progression of the integral process signature, again showing comparatively low values as the print proceeds.  Finally, the bottom panel reveals average maximum temperature versus layer number and again evidences a trend towards lower process temperatures as printing proceeded.}
\label{fig:Fig6}
\end{figure}

Moreover, as highlighted in the introduction, the LPBF process is extremely sensitive to the boundary conditions of the fusion step and these changes are visible in the process signatures compiled in Fig.~\ref{fig:Fig6}.  The top panel, derived from the micro-CT data, shows that density typical of quality LPBF is achieved for about the first $100$~layers.  After layer 100, where the laser powder is reduced to $90$~W, significant porosity results; moreover, the stability of the process does not return to its prior equilibrium when the laser power is returned to $100$~W at layer $150$.  As the density is also observed to decline in this range, as far as $\approx 92$\% in layer 175, this may imply incipient print failure. Considering all cooling events observed by any spectrometer channel, Fig.~\ref{fig:Fig6}b shows the average cooling rate of the part decreases from around $0.3$~ms$^{-1}$ to $0.2$~ms$^{-1}$ over approximately the first 40 layers, where thermal energy is rapidly dissipated into the cool build plate.  

As the print progresses, thermal conduction to the build plate is decreased and thermal energy begins to accumulate in the part and build plate.  Considerable, yet stable, variation is observed in the cooling rate signature through layer~100, including a cyclical pattern that is hypothesized to result from interaction of the meltpool plume with the process laser (i.e., a process disturbance when the laser scan direction is approximately in opposition to the direction of gas flow).  From layer~100, a rough pattern of lower highs is observed and coincides with layers having low density.

Likewise, the next panel down shows that the time-temperature integral climbs over the initial $40$~layers, indicating that, on average, the meltpool is surrounded by increasingly hot material for a longer duration.  This metric trends downward from about layer~$80$ and stabilizes at a relatively low value around layer~$125$.  Finally, average maximum temperature is also plotted in the bottom panel.  Maximum temperatures at the beginning of the print are in excess of $2000$~K; this trend decreases to a value just over the liquidus point ($1739$~K) near layer $100$ and remain at that level for the duration of the print.  In some layers, the maximum temperature appears insufficient to achieve melting.  In combination, it appears that the scan parameters initially generate near full density where maximum temperatures and cooling rate are high.  This suggests a small meltpool that is hot enough to completely melt the feedstock and a portion of the previously fused material.  As the build progresses, the combination of low cooling rate and low maximum temperatures near liquidus are hypothesized to indicate large areas of the part becoming simultaneously molten (i.e., insufficient scan speed to maintain desirable meltpool dimensions) and this is consistent with the balling defects observed in the micro-CT data.  It is interesting to note that none of the process signatures, nor the density, return to their prior level at layer~$150$ where the laser power is returned to $100$~W.  Again, it appears that the process has become so irregular as to prevent a return to the process window consistent with achieving full density.

\subsection{Porosity Detection}

While examination of spatial and temporal temperature transients gives qualitative insight into component quality, we now correlate specific process signatures to the presence of voids in the test artifact identified by micro-CT. For this, the volume directly sensed by each spectrometer channel during fusion of a layer in the test artifact is taken to be a cylinder with a volume equal to the product of the area observed by each channel (fiber) and layer thickness. In other words, each of the $3800$ time series is associated with a $92$~\textmu m diameter, $30$~\textmu m tall cylindrical volume in the test artifact. These corresponding cylindrical volumes within the micro-CT dataset are found by a direct accounting of known and measured shifts, scales, and rotations.  As described above, process signatures are extracted for each of these cylinders and are thresholded.  Process signature values outside of the threshold are taken to indicate porosity in the cylinder and are called alarms.  

The severity of porosity within each cylinder is quantified by summing the number of porous voxels and reporting the diameter of a sphere with an equivalent volume.  Ambiguity is reduced by this convention where, if a plurality of pores exist within an alarm cylinder, it is impossible to say which pores (individually or in aggregate) contributed to an alarm with the present level of experimental sophistication.  A void in the CT data is taken to be detected if it lies in an alarm cylinder in the thresholded process signature data, otherwise it is missed.  Conversely, an alarm is positive if porosity is measured in the cylinder and is otherwise false.  Finally, it is common that the location of an alarm is vertically displaced with respect to the pore that manifests it (e.g., abnormally high temperatures can result from fusing material above a pore that impedes thermal conduction into previously solidified material).  To consider this phenomenon, alarms are given a deviation allowance of $\pm 1, \, 2, \, \mathrm{or} \, 3$~layers, artificially making the alarm cover a taller region in the test artifact, prior to considering detection probabilities in the following analysis.

It should be noted that this is effectively a worst-case assumption.  For example, two small pores in a cylinder that are individually of insignificant size are considered to be indistinguishable from one large, volume-equivalent pore.  If a detection is claimed, even if the small pores would have been detected individually, credit is taken only for detecting the volume-equivalent larger pore.  If a detection is missed in this example, it is counted as having missed the volume-equivalent pore which is also worse than the physical situation.  An additional complication is that each channel is hypothesized to be sensitive to pores that lie in a larger volume that surrounds the volume nominally sensed.  As the dimensions of this volume-of-sensitivity are presently unknown, no lateral allowance is made.  Thus, it is also possible that a fraction of alarms considered to be false in the present analysis are actually valid, and indicate porosity just outside of the directly-observed volume.

\begin{figure}[htbp]
\begin{center}
	{
	\includegraphics[trim = {0.95in 2.4in 5.1in 1.55in}, clip, scale=1, keepaspectratio=true]{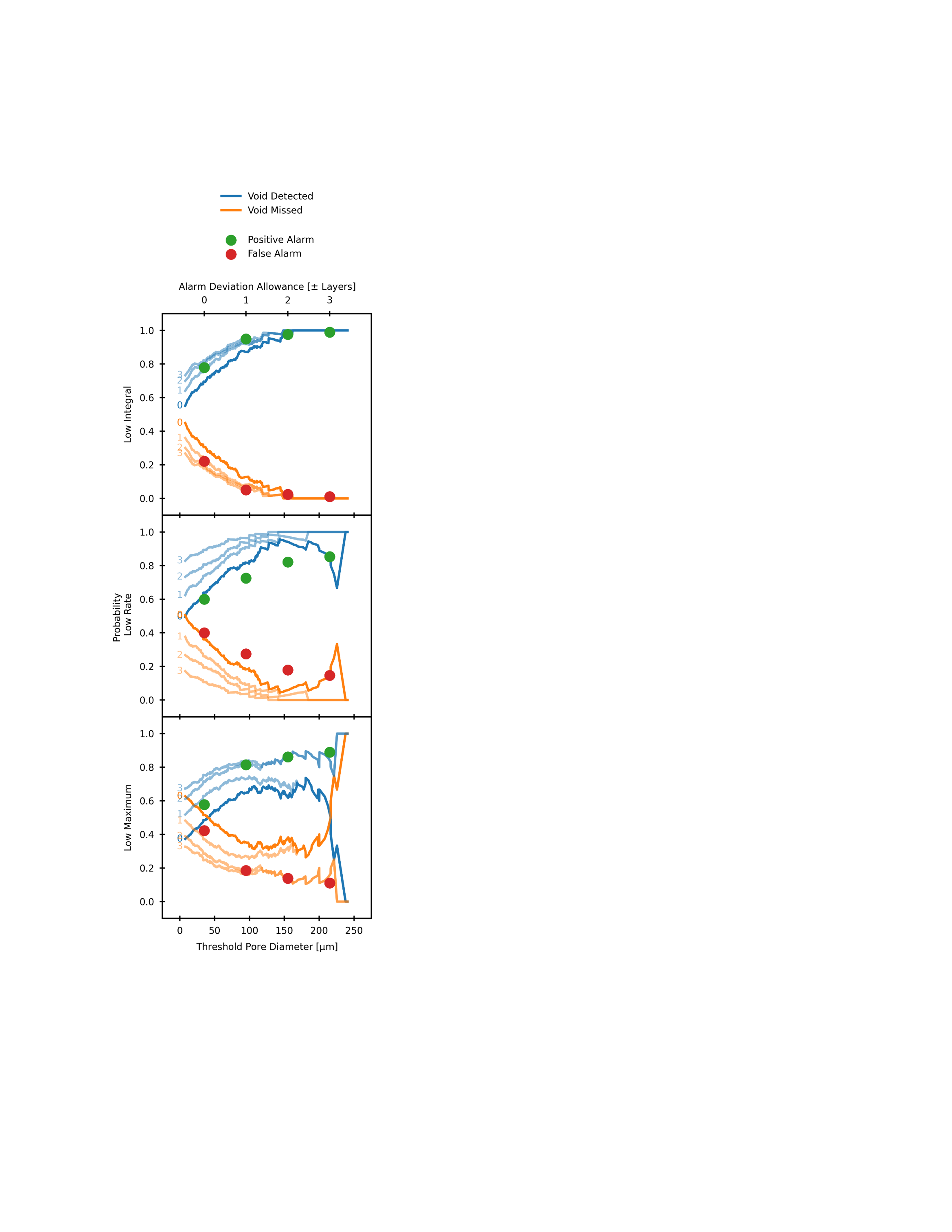}
	}
\end{center}
\caption{Pore detection probabilities for a selection of process signatures.  In the top panel, low integral is used as a process signature.  The bold lines correspond to detection probabilities without a vertical allowance.  Allowing an increasingly large vertical allowance of $\pm 1, \, 2, \, \mathrm{or} \, 3$~layers always improves void detection probability, represented by the faint lines.  This also improves the rate of positive versus false alarms, as represented by the dots.  The center and bottom panels are identical, albeit using low cooling rate and low maximum temperature as process signatures, respectively.}
\label{fig:Fig7}
\end{figure}

Upon this basis, Fig.~\ref{fig:Fig7} presents cumulative detection probabilities for all pores of a certain diameter or larger, using process signatures of low integral, low cooling rate, and low maximum temperature.  Low integral is taken to indicate porosity by one of two mechanisms.  In one case, material that absorbed insufficient energy to fully melt may have low maximum temperatures, reducing this process signature.  Alternatively, and perhaps more relevant to the nature of porosity in the test artifact, low integral may result from rapid dewetting of an area when a balling effect occurs (i.e., high temperatures are maintained only for a short period of time).  For this signature $\approx 50$\% of porosity greater than $4.3$~\textmu m is detected with no vertical allowance. However, performance is greatly improved with a $\pm 1$~layer deviation allowance, after which 64\% of all pores and 71\% of pores greater than $20$~\textmu m are resolved with a 5\% probability that an alarm is false.  

The center panel of Fig~\ref{fig:Fig7} presents an identical analysis of low cooling rate, indicating underlying porosity or large balls of slowly-cooling material.  Setting the threshold for roughly equal rates of pore detection, this signature shows a comparatively higher rate of false alarms.  However, that the shape of the detection probability versus pore size curves differs with respect to the low integral signature.  For example, this process signature appears more sensitive to smaller pores than low integral.  Finally, the bottom panel of the figure shows that low maximum temperature exhibits a low probability of detection, particularly of intermediate and large sized pores $\geq 100$~\textmu m in diameter.

%% file: conclusion.tex
\section{Discussion}

Refinements to this implementation are the subject of ongoing development, including expanding the size of the components that can be printed with this technology.  Instruments based upon Dyson~\cite{Dyson1959} or Offner~\cite{Offner1975} relays are suggested for high-performance designs~\cite{Cobb2006,Lobb1997,Mouroulis2018}, and systems following Schwarzschild's work on mirror telescopes~\cite{Schwarzschild1905} are also possible at a higher level of complexity~\cite{Mouriz2011}.  Regardless of the topology selected, one of the ultimate limits to the field of view of such an instrument is the number of pixels in the sensor, which must be at least the number of wavelengths times the number of channels.  Future study must determine the minimum number of wavelengths necessary for robust TES in this application; yet, even if the current number is necessary, there theoretically exists room for more than 10,000 channels on the sensor used here.  The other critical limitation is the field of sensitivity of each channel.  This remains uncertain for the existing instrument, as are the factors that most strongly influence it.  Nevertheless, a $10\times10$~mm$^2$ field of view is possible using the combining the current fiber spacing and detector size into a lower bound, although we hypothesize that a larger area can be observed by increasing the fiber spacing without compromising on defect detection probability.

Likewise, there are many promising directions for improving the temperature-emissivity separation element of this work.  At a simple level, both regularization at the estimation step and filtering could be used to reduce the influence of measurement noise.  More advanced methods, particularly subspace methods~\cite{Adams2006,Acito2019}, for performing TES may also improve temperature fidelity and even make emissivity available as a process signature in its own right.  Among other options in this direction, vertex component analysis (VCA) is popular in view of its simplicity~\cite{Nascimento2005_VCA, Barberis2013}. Alternatively, SMACC (sequential maximum angle convex cone) is purpose-designed to handle datasets in which observed emissivities are highly correlated~\cite{Gruninger2004}.  This appears to be an excellent fit to the problem at hand, especially if using emissivity as an LPBF process signature, where there are a limited number of material states that are of interest and the differences in spectral emissivity indicating defect-prone process conditions may be subtle.

\section{Conclusion}

Temperature-emissivity separation is capable of retrieving accurate temperature measurements during material fusion in the LPBF process, despite great inherent uncertainty of the spectral emissivity of powdered, molten, and solid build material.  Construction and integration of an imaging spectrometer with an LPBF testbed enables this experimental campaign, and is fully described here as a reference design.  The joint performance of this instrument and analysis method is quantified using point melting experiments that authenticate temperatures to $\pm 28$~K over a temperature range of $633$ to $1940$~K, soundly surpassing the capabilities of typical wideband and two-color pyrometry techniques.  Operability is then confirmed by observing the fabrication a 316~SS LPBF test artifact, where fully-dense regions show remelting at temperatures well above liquidus and rapid cooling.  Fusion boundary conditions are also studied with this dataset, where mean process temperatures climb and cooling rate falls as the print progresses.  Finally, an exploratory investigation correlates high-fidelity thermal process signatures to porosity in the test artifact.  The integral process signature, quantifying time-at-temperature, is predictive of 71\% of pores greater than $20$~\textmu m with a 5\% false alarm rate if a minimal allowance is made for the relative vertical location of pores and their corresponding alarms.  These results establish efficacy of TES-derived process signatures in quality control and they are equally applicable to other aspects of the LPBF workflow.  Deterministic process tuning is made possible.  For example, scan parameters can be precisely updated from part to part in response to measured process conditions to avoid defect-prone fusion conditions.  Similarly, closed-loop process control is also expected to benefit from improved temperature fidelity at a higher level of complexity.

%% file: supplement.tex
\beginsupplement
\pagebreak
\section{Spectrometer}
\label{sec:SupSpectrometer}

\begin{figure*}[htbp]
\begin{center}
	{
	\includegraphics[trim = {0.8in 4.3in 1in 1.8in}, clip, scale=1, keepaspectratio=true]{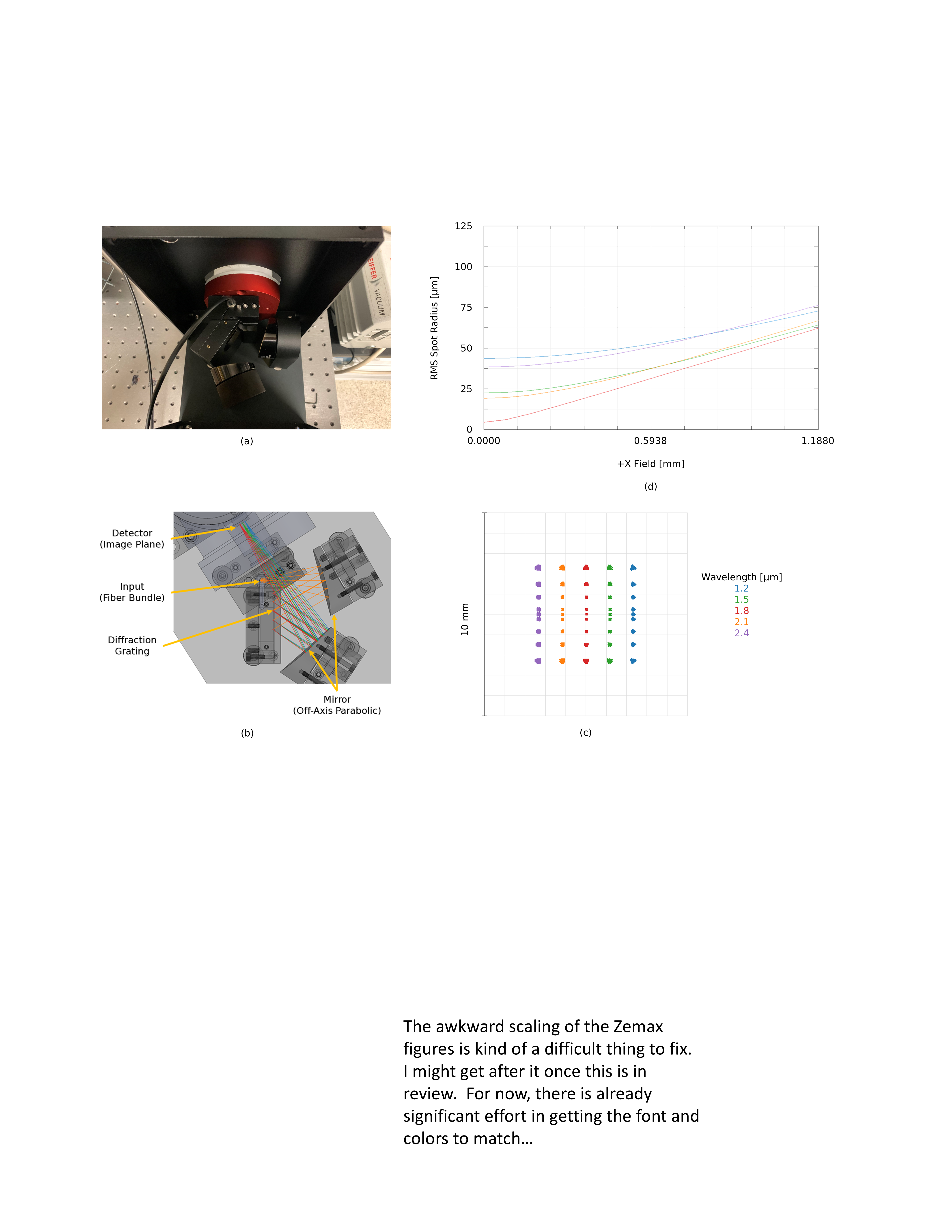}
	}
\end{center}
\caption{(a) Optical path of the as-built imaging spectrometer.  (b) Overlaid spectrometer solid model and ray trace, with critical components labeled. (c) Full field spot diagram for the spectrometer design.  (d) RMS spectrometer spot diameters as a function of field height for the central wavelength.}
\label{fig:FigS1}
\end{figure*}

The imaging spectrometer originally described in~\S~\ref{sec:MethodsSpectrometer}, is purpose-built for this study.  To explain the optical model in Table~\ref{table:SpectrometerPrescription} and the CAD rendering in Fig.~\ref{fig:FigS1}b, the spectrometer-end of the fiber bundle enters the instrument vertically.  Light from the bundle is first redirected by a fold mirror (Surface 2, Edmund Optics PN 48-385), then collimated by a first off-axis parabolic mirror (Surface 7, Edmund Optics PN 35-622) with a focal length of $54.4$~mm.  The diffraction grating (Newport Optics / Richardson Gratings PN 53006BK02-606R) is then placed at a convenient distance from the first mirror.  This optic is a replicated grating on an N-BK7 substrate.  It features a groove density of $30$~grooves/mm and has an unprotected gold coating.  It is nominally blazed for $1.10$~\textmu m light (or with a $1.20^\circ$ angle) in the Littrow configuration; following~\cite{Newport2005} to account for the present geometry, it is most efficient at $1.22$~\textmu m.  A second off-axis parabolic mirror (Surface 12, Edmund Optics PN 35-641) with a focal length of $108.9$~mm  is again placed a comfortable distance away, in the direction of the optical path and where this path is tracked with a chief-ray solve.  The final surfaces before the image plane (14, 15, and 17) correspond to components of the camera dewar (IRCameras IRC806HS): the dewar front, to track and prevent mechanical interference; the dewar window; and the f/2.3 cold stop.  Two compensators are permitted in analysis of the the optical design and implemented in the mechanical design.  Specifically, the distance between the end of the fiber bundle and first mirror is adjustable via a flexure clamp (visible from the top in Fig.~\ref{fig:FigS1}a) that secures the fiber bundle ferrule to ensure light is collimated before reaching the diffraction grating.  The distance between the second mirror and detector is also adjustable with the assistance of a micrometer drive (Mitutoyo PN 148-811) to ensure light is focused at the detector.

Expected performance of the instrument is illustrated via Figs.~\ref{fig:FigS1}c and~\ref{fig:FigS1}d that provide a full field spot diagram and plot RMS spot diameters, respectively.  To explain, the former represents the spots on the detector as a function of field location (here, taken to be a series of locations along the linear end of the spectrometer bundle) in the vertical direction and a selection of wavelengths in the horizontal direction.  The latter plots the RMS field diameter as a function of field location for the same wavelengths.  As is characteristic of parabolic mirrors, performance is extremely good at the center of the image yet coma and astigmatism become substantial off axis.

Two related notes are made concerning the spectral and temporal resolution of this instrument.  First, the length of spectra on the detector is readily estimated from the provided data as $4.387$~mm or $220$~px using the $20$~\textmu m pixel pitch, if the fiber diameter is neglected.  More precise values of $4.762$~mm and $238$~px are found via Zemax ray trace and suggests a dispersion of roughly $5\, \mathrm{nm/px}$.  As the camera detector is $640$~px wide, this selection provides the option to increase its sample rate by a factor of~$2$ if windowing is applied in this direction.  An additional factor of~$2$ improvement in temporal resolution is possible by windowing in the vertical direction.  Second, as a slit is not employed, the finite diameter of the fibers in the fiber bundle is the principle limitation to spectral resolution and imaging quality is a secondary factor.   For combinations of field and wavelength that appear near the center of the detector, the monochromatic image of a fiber is a little under $8$~px wide in the dispersion direction at FWHM and aberration is negligible.  Measurement with a filter (Thorlabs FBH1400-12) suggests a value of $8.15$~px after accounting for the filter passband.  As such, the spectra are rebinned into $238/8 \approx 30$ bands.  Imaging performance increasingly degrades for fields and wavelengths that land more distantly from the center of the detector and therefore the spectra also become increasingly, if only slightly, oversampled.

A complete instrument calibration to radiance units is performed.  For this purpose, a cavity-type blackbody simulator is used, comprising a mechanically reconfigured Pentron JP1200 porcelain furnace with a cast zirconia plug.  A $5$~mm aperture in the plug provides optical access to the furnace interior.  The complete optical path (LPBF testbed environmental window, ADM lens, ADM lens interface, fiber bundle, and spectrometer) is replicated on a test stand, and spectra are recorded for a selection of $9$ temperatures spanning $1023$ to $1423$~K.  Theoretical radiance is computed per Eq.~\ref{eq:SpectralPlancksLaw} for each temperature, and a single correction function is fit that relates detector-counts per second to radiance units of W/m$^2$/sr/\textmu m.

\begin{landscape}
\begin{table}[bt]
\centering
\tiny
\caption[Spectrometer optical prescription.]{Spectrometer optical prescription.  Under surface type, "Coordinate" is short for Coordinate Break and "Diffraction" for Diffraction Grating.  Part numbers given under the comment field are from Edmund Optics, except for the diffraction grating from Newport.  All mirror surfaces are unprotected gold except P/N 48-385 which is protected gold.}
\label{table:SpectrometerPrescription}
\begin{tabular}{@{}*{21}c@{}}
  
  \toprule[1.5pt]
  \multicolumn{2}{c}{\head{Surface}} &
  \multicolumn{1}{c}{\head{Comment}} &
  \multicolumn{1}{c}{\head{Radius}} &
  \multicolumn{2}{c}{\head{Thickness}}&
  \multicolumn{1}{c}{\head{Material}}&
  \multicolumn{2}{c}{\head{Semi-Dia.}}&
  \multicolumn{1}{c}{\head{Conic}}&
  \multicolumn{2}{c}{\head{Decenter X}}&
  \multicolumn{2}{c}{\head{Decenter Y}}&
  \multicolumn{2}{c}{\head{Tilt X}}&
  \multicolumn{2}{c}{\head{Tilt Y}}&
  \multicolumn{2}{c}{\head{Tilt Z}}&
  \multicolumn{1}{c}{\head{Order}}\\
  
  \cmidrule{1-21}
  \head{Number} & \head{Type} & & \head{Lines/\textmu m} & \head{Diffract Order}\\
  \cmidrule{1-21}
  OBJ & Standard    &            & Infinity & 0.000   &   &          & 2.090  &   &  0.000 &       &   &         &   &         &   &         &   &       &   &   \\
  1   & Standard    &            & Infinity & 5.000   &   &          & 5.000  & U &  0.000 &       &   &         &   &         &   &         &   &       &   &   \\
  2   & Coordinate  &            &          & 0.000   &   &          & 0.000  &   &  0.000 & 0.000 &   & 0.000   &   & 0.000   &   & -45.000 &   & 0.000 &   & 0 \\
  3   & Standard    & 48-385     & Infinity & 0.000   &   & MIRROR   & 5.875  &   &  0.000 &       &   &         &   &         &   &         &   &       &   &   \\
  4   & Coordinate  &            &          & -15.000 &   &          & 0.000  &   &  0.000 & 0.000 &   & 0.000   &   & 0.000   &   & -45.000 &   & 0.000 &   & 0 \\
  5   & Coordinate  &            &          & 20.000  &   &          & 0.000  &   &  0.000 & 0.000 &   & 0.000   &   & 0.000   &   & 0.000   &   & 0.000 &   & 0 \\
  6   & Coordinate  &            &          & -54.450 &   &          & 0.000  &   &  0.000 & 0.000 &   & 0.000   &   & -30.000 &   & 0.000   &   & 0.000 &   & 0 \\
  7   & Standard    & 35-622     & 101.6    & 0.000   &   & MIRROR   & 39.540 &   & -1.000 &       &   &         &   &         &   &         &   &       &   &   \\
  8   & Coordinate  &            &          & 30.000  &   &          & 0.000  &   &  0.000 & 0.000 &   & 0.000   &   & 0.000   &   & 0.000   &   & 0.000 &   & 0 \\
  9   & Coordinate  &            &          & 0.000   &   &          & 0.000  &   &  0.000 & 0.000 &   & 48.727  & C & 0.000   &   & 0.000   &   & 0.000 &   & 0 \\
  STO & Diffraction & 53-606R    & Infinity & 1       &   & MIRROR   & 14.444 &   &  0.000 & 0.030 &   & 1.000   &   &         &   &         &   &       &   &   \\
  11  & Coordinate  &            &          & -75.000 &   &          & 0.000  &   &  0.000 & 0.000 &   & -52.000 &   & 33.517  &   & 0.000   &   & 0.000 &   & 0 \\
  12  & Standard    & 35-641     & 203.200  & 74.719  & V & MIRROR   & 56.273 &   & -1.000 &       &   &         &   &         &   &         &   &       &   &   \\
  13  & Coordinate  &            &          & 0.000   &   &          & 0.000  &   &  0.000 & 0.000 &   & 12.552  & C & 23.895  & C & 0.000   &   & 0.000 &   & 0 \\
  14  & Standard    & Dewar      & Infinity & 0.000   &   &          & 46.101 & U &  0.000 &       &   &         &   &         &   &         &   &       &   &   \\
  15  & Standard    & Window     & Infinity & 1.000   &   & SAPPHIRE & 7.326  &   &  0.000 &       &   &         &   &         &   &         &   &       &   &   \\
  16  & Standard    &            & Infinity & 5.029   &   &          & 7.279  &   &  0.000 &       &   &         &   &         &   &         &   &       &   &   \\
  17  & Standard    & Cold Stop  & Infinity & 25.273  &   &          & 5.525  & U &  0.000 &       &   &         &   &         &   &         &   &       &   &   \\
  IMA & Standard    &            & Infinity & -       &   &          & 4.845  & U &  0.000 &       &   &         &   &         &   &         &   &       &   &   \\
  \bottomrule[1.5pt]
\end{tabular}
\end{table}
\end{landscape}

\clearpage
\onecolumn
\section{TES Method Comparison}
\label{sec:SupTES}

\begin{figure*}[b]
\begin{center}
	{
	\includegraphics[trim = {0.5in 1.35in 3.6in 1.2in}, clip, scale=1, keepaspectratio=true]{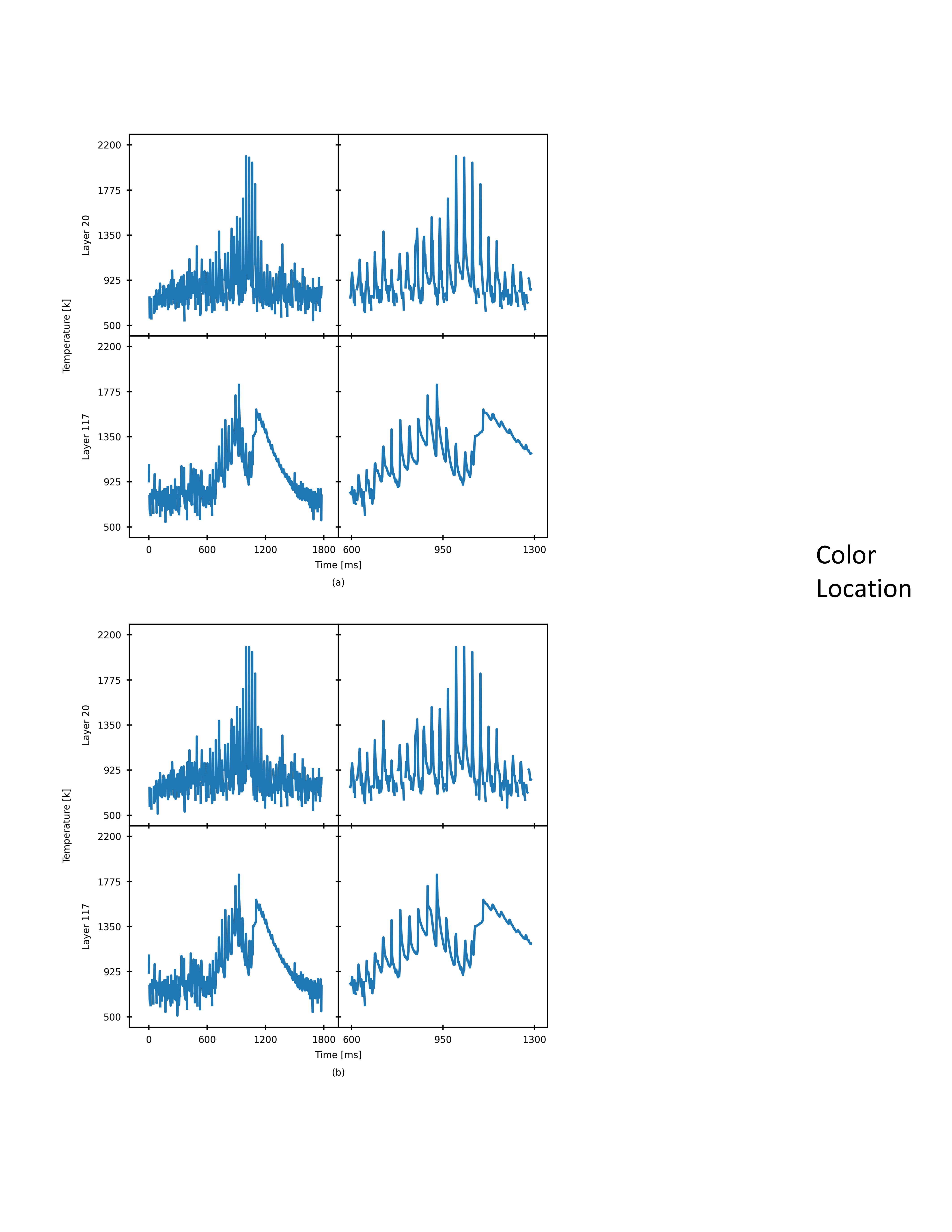}
	}
\end{center}
\caption{Temperatures recovered using the two color (a) and two location (b) TES methods.  The significance of each panel corresponds to those of Fig.~\ref{fig:Fig5}a; direct comparison shows nearly indistinguishable performance.}
\label{fig:FigS2}
\end{figure*}